\documentclass[journal]{IEEEtran}
\usepackage{ifpdf}
\usepackage{cite}
\ifCLASSINFOpdf
   \usepackage[pdftex]{graphicx}
\else
   \usepackage[dvips]{graphicx}
\fi

\usepackage{booktabs}
\usepackage{multirow}
\usepackage{multicol}

\usepackage[cmex10]{amsmath}
\usepackage{algorithm}

\usepackage{algorithmic}
\usepackage{array}
\ifCLASSOPTIONcompsoc
  \usepackage[caption=false,font=normalsize,labelfont=sf,textfont=sf]{subfig}
\else
  \usepackage[caption=false,font=footnotesize]{subfig}
\fi

\begin{document}
%
% paper title
% Titles are generally capitalized except for words such as a, an, and, as,
% at, but, by, for, in, nor, of, on, or, the, to and up, which are usually
% not capitalized unless they are the first or last word of the title.
% Linebreaks \\ can be used within to get better formatting as desired.
% Do not put math or special symbols in the title.
% \title{P2P Software Model To Achieve High Availability In Coronavirus Sampling For COVID-19}
\title{Software in P2P way: a software model without central software and enabling any software to join or leave freely}
%
%
% author names and IEEE memberships
% note positions of commas and nonbreaking spaces ( ~ ) LaTeX will not break
% a structure at a ~ so this keeps an author's name from being broken across
% two lines.
% use \thanks{} to gain access to the first footnote area
% a separate \thanks must be used for each paragraph as LaTeX2e's \thanks
% was not built to handle multiple paragraphs
%

\author{Hong~Su% <-this % stops a space
\IEEEcompsocitemizethanks{\IEEEcompsocthanksitem H. Su is with School of Computer Science, Chengdu University of Information Technology, Chengdu, China. E-mail: suguest@126.com. \\
% \IEEEcompsocthanksitem B. Guo is with the College of Computer Science, Sichuan University, Chengdu, China. E-mail: guobing@scu.edu.cn.
% \IEEEcompsocthanksitem H. Xu is with the School of Intelligence Technology, Geely University of China, Chengdu, China. E-mail: 842466579@qq.com\\
% note need leading \protect in front of \\ to get a newline within \thanks as
% \\ is fragile and will error, could use \hfil\break instead.
}% <-this % stops an unwanted space
\thanks{}}

% The paper headers
\markboth{IEEE LaTeX Version,~Vol.~X, No.~X, X~X}%
{Shell \MakeLowercase{\textit{et al.}}: Bare Demo of IEEEtran.cls
for Journals}

% make the title area
\maketitle

% As a general rule, do not put math, special symbols or citations
% in the abstract or keywords.
\begin{abstract}
  The P2P model encompasses a network of equal peers, whether in hardware or software, operating autonomously without central control, allowing individual peer failure while ensuring high availability. 
  Nevertheless, current P2P technologies primarily focus on hardware-level resilience, often referred to as P2P networks, which do not safeguard against software failures. 
  This paper introduces a pioneering Peer-to-Peer (P2P) software model aimed at enhancing software-level high availability. Diverging from prevalent hardware-centric P2P technologies, this model accentuates the decentralized nature of various software components, or "software peers," which function independently, enabling seamless network entry and exit without relying on central software. 
  The model's collaborative approach cultivates a network topology with multiple autonomous processing paths, ensuring continuous operation through dynamic task allocation in a distributed manner. By surpassing the limitations of traditional redundancy methods, this P2P model provides an adaptive and scalable solution for achieving robust availability. 
  Validation results underscore the model's effectiveness in enhancing the probabilities of successful task processing while ensuring high availability.
  
\end{abstract}

% Note that keywords are not normally used for peerreview papers.

% For peer review papers, you can put extra information on the cover
% page as needed:
% \ifCLASSOPTIONpeerreview
% \begin{center} \bfseries EDICS Category: 3-BBND \end{center}
% \fi
%
% For peerreview papers, this IEEEtran command inserts a page break and
% creates the second title. It will be ignored for other modes.
\IEEEpeerreviewmaketitle

\section{Introduction}
Ensuring high availability is critical for the stability and functionality of any system, particularly in domains where system reliability directly impacts critical operations. Industries such as finance, healthcare, and telecommunications rely heavily on uninterrupted services, emphasizing the pivotal role of high availability in sustaining business continuity and enhancing the user experience \cite{VayghanLA}. Recent studies highlighting the increased reliance on cloud-based infrastructures further emphasize the need for robust high availability strategies to prevent service disruptions and data loss, ultimately safeguarding against revenue loss and reputational damage \cite{LiC}. By implementing comprehensive high availability measures, organizations can effectively mitigate the risks associated with system downtime and ensure seamless, uninterrupted operations, thereby strengthening their competitive edge in a dynamic and interconnected marketplace.

Peer-to-Peer (P2P) networks, known for their decentralized structure and resilience to individual peer failures, have primarily focused on hardware resilience, neglecting the critical aspect of software-level high availability. Although hardware-centric P2P technologies have facilitated resilient networking, the prevalence of software vulnerabilities and failures in modern computing environments remains a persistent challenge. As reliance on software systems continues to grow, the necessity for a comprehensive software-level high availability framework within P2P networks becomes increasingly evident. 

While such architectures might leverage distributed hardware environments like clouds or P2P setups,  their capacity to avert system outages is largely limited to hardware failures within specific hardware subsets, rather than accounting for potential vulnerabilities in the software domain.  
The current software model is characterized by: (1) employing a singular software for task handling, or (2) utilizing multiple software components coordinated by a specific software, or (3) incorporating multiple software pieces featuring identical or akin binary code. This architectural framework inherently fosters singularities at the software level. Additionally, some backup software might be employed (similar to N-version programming), but this entails a fixed count of software instances, precluding the freedom for software to seamlessly join or depart. 

To address these issues, this paper introduces a Peer-to-Peer (P2P) software model. The P2P system comprises a set of software that collaboratively establishes a network architecture devoid of singular task processing pathways, enabling seamless software replacement. Termed \textbf{software peers}, these softwares possess the characteristics of independent operation, freedom from central coordination, and autonomous development. Within the P2P system, multiple processing paths ensure task continuity even in the face of the failures of some paths. 

This model brings forth two key advantages: (1) The availability of multiple processing paths eliminates singularities, enhancing task processing reliability. These paths, formed by different softwares that are independently designed, coded, and verified, forge a network structure that amplifies task success probabilities. (2) The model facilitates the integration of new software alongside existing counterparts, enabling a smoother task transition between them upon failures, resulting in heightened availability.

This paper's contributions are threefold:
\begin{enumerate}
    \item We propose the P2P software model, composed of independent softwares that operate without central coordination, with each software functioning as a self-reliant \textbf{software peer}. These software peers offer distinct processing paths, with task switching mechanisms activated upon failures. The predecessor software handles path switching, eliminating the need for central control and enhancing path redundancy.
    \item We introduce a layered P2P model, enabling finer-grained interactions among software. This approach, partitioning software into layers, promotes task switching within layers rather than software shifts, leading to augmented processing paths and network topology. This stratified structure enhances task processing efficiency, success rates, and adaptability to intricate tasks.
    \item We propose a method to evaluate P2P model availability through topology analysis. By representing the P2P model as a graph elucidating interactions among software peers, we identify coupled paths and optimize performance metrics such as task time and success probability. This topology-based approach offers insights into path relationships, enhancing overall model availability.
\end{enumerate}

The remainder of this paper is organized as follows. Section II presents related work. Sections III and IV describe the P2P software model and P2P topology, respectively. Section IV describes the validation results, and Section V concludes the paper.

\section{Related Work}
The central objective of this paper is to enhance system availability by leveraging the peer-to-peer (P2P) software model, establishing independent redundancy at the software level and facilitating seamless transitions to alternate processing paths. To provide context, we begin with an overview of pertinent research on redundancy, focusing on hardware and P2P technology. Subsequently N-version programming and conduct a comparative assessment are explored. Additionally, alternative strategies pursued to achieve heightened availability are discussed.

\subsection{P2P Technologies}
P2P technology, predominantly employed for hardware-level redundancy, distributes tasks across peer nodes \cite{TseliosC}, aiming to obviate reliance on centralized nodes \cite{MarzalS}. These nodes, which can function as servers or clients \cite{HuynhTK}, ensure continuity of functionality when one node falters \cite{BauwensM}. P2P technology also facilitates load distribution and resource sharing \cite{SinghVP}, yet the current approach pertains to \textbf{hardware-level} peer-to-peer, with corresponding software retaining centralized features. For instance, a centralized software may coordinate others (e.g., a task scheduler in grid computing) or software instances may feature identical or similar binary codes (e.g., BitTorrent \cite{SoderbergE}).

P2P research centers on network structure, efficiency, and overarching frameworks. (1) Effective P2P network formation is explored through unstructured and structured networks \cite{StutzbachD} \cite{CrainiceanuA}. (2) Efficiency and scalability are addressed, with work on effort-based incentives \cite{RahmanR}. (3) General frameworks, such as middlewares, foster P2P software development \cite{ShudoK}. P2P computing, a key application, assigns tasks to multiple servers, where specific nodes (supernodes) function as intermediaries \cite{JMishra}. Notably, the decentralized nature of P2P computing is marred by super nodes' centralization, as their failure can trigger system-wide outages. P2P technology has also extended to software level applications \cite{SuhFDAM}, yet independent redundancy and software relationship topologies remain unexplored.

An important application of P2P technology is P2P computing, in which tasks are assigned to multiple servers to execute computations that require a large amount of resources. During this process, specific nodes (supernodes) act as intermediaries between users and the task-running server \cite{JMishra}. The super node has the characteristics of centralization: because if it fails, the whole system will fail. P2P computing belongs to grid computing. Grid computing is a kind of distributed computing, which has a center to schedule tasks and collect final results\cite{ZhangS}.

With the development of P2P technology, the P2P method at the software level is proposed. Work\cite{SuhFDAM} proposes P2P software in the blockchain environment. This work aims to achieve a fully decentralized application model, that is, P2P software on P2P hardware, and how to achieve consistency between P2P software. However, it fails to address the issue of independent redundancy among software, and does not delve into the relationship topology between different software.

\subsection{N-version Programming}
N-Version Programming (NVP) is a software engineering technique aimed at enhancing software reliability and fault tolerance. NVP involves creating multiple independent versions of the same software system using varied methods, algorithms, or compilers. These versions should yield identical outputs for a given input under normal conditions, enabling error detection and mitigation through output comparison \cite{LevitinG} \cite{GirdharV}.

NVP and the P2P model represent distinct approaches to bolstering software reliability and fault tolerance. (1) NVP encompasses multiple independently developed software versions, aiming to detect errors via output differences. However, NVP is constrained by a fixed number of versions, limiting adaptability and scalability in dynamic environments, and introducing complexities associated with consensus mechanisms and common-mode errors. (2) In contrast, the P2P model emphasizes high availability through different software units executing tasks via diverse paths, forming a network topology with redundancy. Unlike NVP, the P2P model permits flexible addition or removal of software, enhancing adaptability to changing requirements. The decentralized execution and failover mechanisms in the P2P model differentiate it from NVP's error detection via output comparison.

\subsection{Other Methods to Achieve High Availability}
Load balancing techniques, such as Static Load Balancing (SLB) and Dynamic Load Balancing (DLB), distribute tasks among servers to prevent overloading \cite{ShafiqDA} \cite{MishraSK}. Both rely on a balancing server, which if compromised, can lead to system failure, and task execution servers typically feature identical binaries, potentially introducing common errors.

Disaster Recovery (DA) focuses on data storage resilience during natural disasters \cite{MendoncaJ}. DA entails storing data copies in different locations, restoring data from unaffected sites when damage occurs. In contrast, the P2P model concentrates on proactive redundancy through independent software units with diverse paths, ensuring high availability and fault tolerance. DA primarily centers on post-event data recovery, while the P2P model emphasizes distributed redundancy and failover mechanisms.

In summary, while P2P technology, N-version programming and other methods offer valuable insights into enhancing software reliability and availability, the proposed P2P software model introduces a unique approach by focusing on independent redundancy at the software level, facilitating dynamic task execution through varied paths.

\section{Peer-to-Peer Software Model (P2P Model)}

\subsection{Motivation}
In a system, both hardware and software components are vulnerable to failures that can lead to system outages. While redundant hardware is often used to address hardware failures, a centralized scheduling mechanism might be necessary. Peer-to-peer (P2P) technology offers an alternative approach to achieving redundancy and reliability in such systems. However, the conventional P2P mechanism operates at the hardware level.

Our proposed P2P model extends the concept of P2P from hardware to software. In this model, software peers can seamlessly redistribute processing tasks, allowing for swift transitions in the event of software failures. To achieve this, two requirements must be met:

(1) Static Independence Requirement ($REQ_1$): Each software has at least one candidate software to be replaced, and the candidate software should be independent of each other. 

(2) Dynamic Independence Requirement ($REQ_2$): Centralized control of processing is avoided in the P2P model due to two crucial reasons: (2a) A failure in the centralized software could lead to a system-wide breakdown, and (2b) the necessity for software replaceability, as no software can be guaranteed to be failure-free. To address these concerns, the model emphasizes the importance of software equality, allowing each piece of software to be considered a peer, promoting a more balanced and robust architecture. 

\subsection{P2P Software}
To fulfill $REQ_1$ and $REQ_2$, a system should consist of multiple independent softwares capable of redistributing processing tasks seamlessly. This ensures that individual software failures don't disrupt the entire system. This concept is encapsulated in the term \textbf{P2P software}, where a set of autonomous softwares collaboratively complete tasks without central coordination. Each individual software in this framework is referred to as a P2P software or a P2P software peer.

The introduction of P2P software necessitates a shift in related terminology. An P2P \textbf{system} or \textbf{application} now comprises multiple software peers that collectively execute tasks. Software peers can be grouped, and each group can independently complete tasks. Thus, a P2P system or application is not a single software, but an set of softwares.

A formal description of a P2P application includes the notation $task.completeBy(soft_a, soft_b,...)$, indicating that task $task$ can be executed by software peers ($soft_a$, $soft_b$,...). Assuming a P2P application $P2PAPP$ aims to complete task $task$, it consists of P2P software peers ranging from $soft_1$ to $soft_n$, as shown in \eqref{app_definition}. An P2P application must satisfy the requirements defined in formulas \eqref{app_division_definition}, \eqref{app_division_independent}, and \eqref{app_division_complete}. These requirements ensure that the software set can be divided into independent groups ($P2PAPP_1$, $P2PAPP_2$, ...) capable of completing the task independently.

\begin{equation} \label{app_definition}
  P2PAPP = \{soft_1, ..., soft_i, ..., soft_n\}
\end{equation}

\begin{equation} \label{app_division_definition}
    P2PAPP = P2PAPP_1 \cup P2PAPP_... \cup P2PAPP_m 
\end{equation}

\begin{equation} \label{app_division_independent}
  \forall P2PAPP_m, P2PAPP_n \rightarrow \ \ P2PAPP_m \cap P2PAPP_n = \emptyset
\end{equation}

\begin{equation} \label{app_division_complete}
    \forall P2PAPP_i \rightarrow \ \ task.completeBy(P2PAPP_i)
\end{equation}

\subsubsection{Working Path} 
\begin{figure}[htb]
      \includegraphics[width=3.5in]{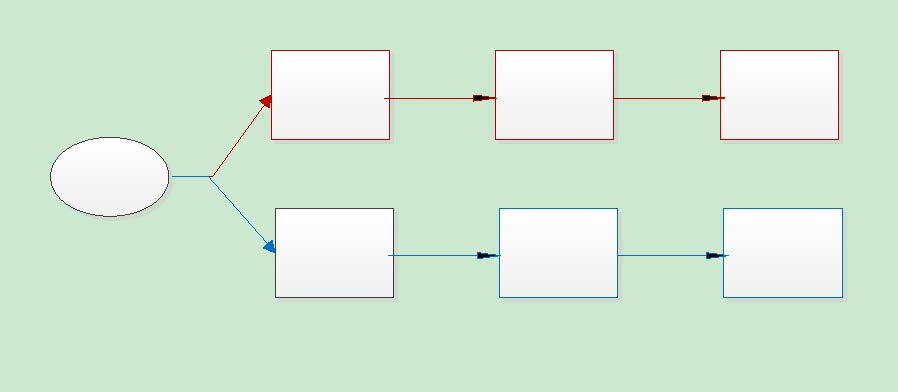}
      \caption{Working path example. The red path and the blue path are two different working paths. }
      \label{workingPath}
\end{figure}

Each group in \eqref{app_division_definition} can complete the target task, which can be regarded as a path in task processing.
A working path represents the sequence of software peers from the initiating terminal to the final processing software for completing a task. 
An example is shown in Figure \ref{workingPath}. A terminal can choose any one of the working paths to complete the task. If a software in one working path fails, task processing can be switched to other working paths.

The number of nodes in the working path corresponds to the number of software in the corresponding group ($P2PAPP_i$). If a single software can complete the job, the working path contains only two nodes - the terminal and the processing software.
However, in some cases, a working path may consist of several software peers that cooperate to accomplish a task, resulting in more than two nodes in the working path.

In order to support task switching to run on other paths, the target task should be \textbf{repeatable task}, that is, a task that can be run multiple times. Since we cannot assume that the task will succeed every time, repeatable execution is necessary if the task is to be allowed to retry on failure. 
In order to identify the same task when it can be executed repeatedly, each task contains a unique identifier, $id$, as shown in \eqref{task_definition}. Meanwhile, repeated execution will also generate multiple results, and the results also contain the identifier of the corresponding task, as shown in \eqref{result_definition}.

\begin{equation} \label{task_definition}
    task: id, ...
\end{equation}

\begin{equation} \label{result_definition}
    Result_{task}: id, ...
\end{equation}

\subsubsection{Function Layers} 
If a working path consists of multiple software peers, each software peer is one layer, and we call it a functional layer.
The layer division of each working path may be the same or different.
Several paths having the same layer means that the corresponding layers have the same interface, including input and output interfaces.
If some software has the same layer division, they form a \textbf{software pool}. These software peers can be candidates for each other, and task processing can switch between them.
As these software peers are independent, the switching may have more chances to be completed successfully.

\begin{figure}[htb]
    \includegraphics[width=3.5in]{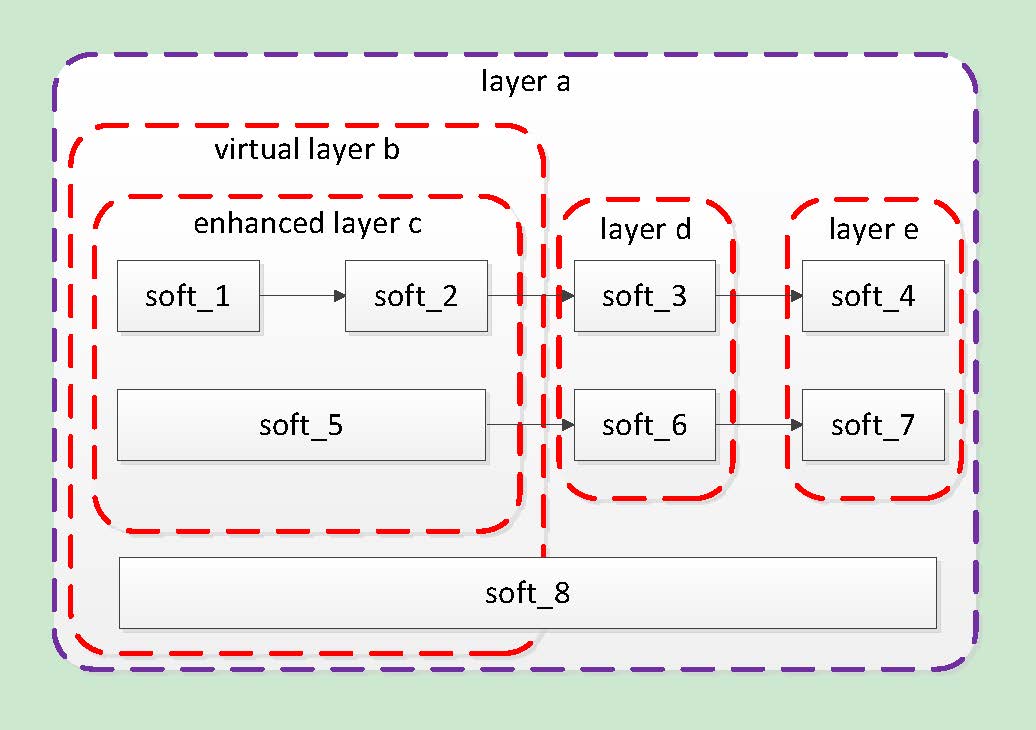}
    \caption{Different layers of P2P software. Layers d and e are normal layers because they have the same input and output interfaces. Layer c is an enhanced layer because $soft_1$ and $soft_2$ can be combined to make the input interface of $soft_1$ and the output of $soft_2$ the same as that of $soft_5$. And layer b is a virtual layer, because $soft_8$ has no corresponding input and output interfaces, but they can have the same processing function.}
    \label{enhanced_layer_demonstration}
\end{figure}

Functional layers can be classified into three types: normal, enhanced, and virtual layers. If working paths have the same layer division, the corresponding layer is called a normal layer. For example, $soft_3$ and $soft_6$ in Figure \ref{enhanced_layer_demonstration} form a normal layer (layer d). In addition to the normal layer division, there are two other layers, the enhanced layer and the virtual layer.

Enhanced layer. If some softwares on one path can be combined to have the same interface as software on the other path, they can be considered a layer, called an enhanced layer. In Figure \ref{enhanced_layer_demonstration} layer c is an enhanced layer because $soft_1$ and $soft_2$ can be combined to have the same input and output as $soft_5$.

Virtual layer. Virtual layers encompass software components with identical processing functions, yet lack a shared interface across distinct processing paths. However, from a logical standpoint, these virtual layers exhibit a uniform interface. In Figure \ref{enhanced_layer_demonstration}, layer b is a virtual layer because $soft_8$ does not have input and output interfaces corresponding to $soft_1$ or $soft_5$, but they can have the same processing functions.

Different layer types affect the amount of software involved when switching from one path to another. For comparison, we introduce \textbf{switch path}. A switch path is a path from faulty software to software that can retry. The length of the switch path $len_{switch}$ is the amount of software along the switch path. The switching process consists of two steps.

(1) When a software peer fails, the task processing returns to the upper layer until there is a software peer that can continue to execute: the software peer has the same function and is not a failed software peer. The path involved in this step is called the back path.

(2) In the second step, the task switches to the new software peer to continue processing. The path to switch to new software peer is called the forward path.

\begin{figure}[htb]
    \includegraphics[width=3.5in]{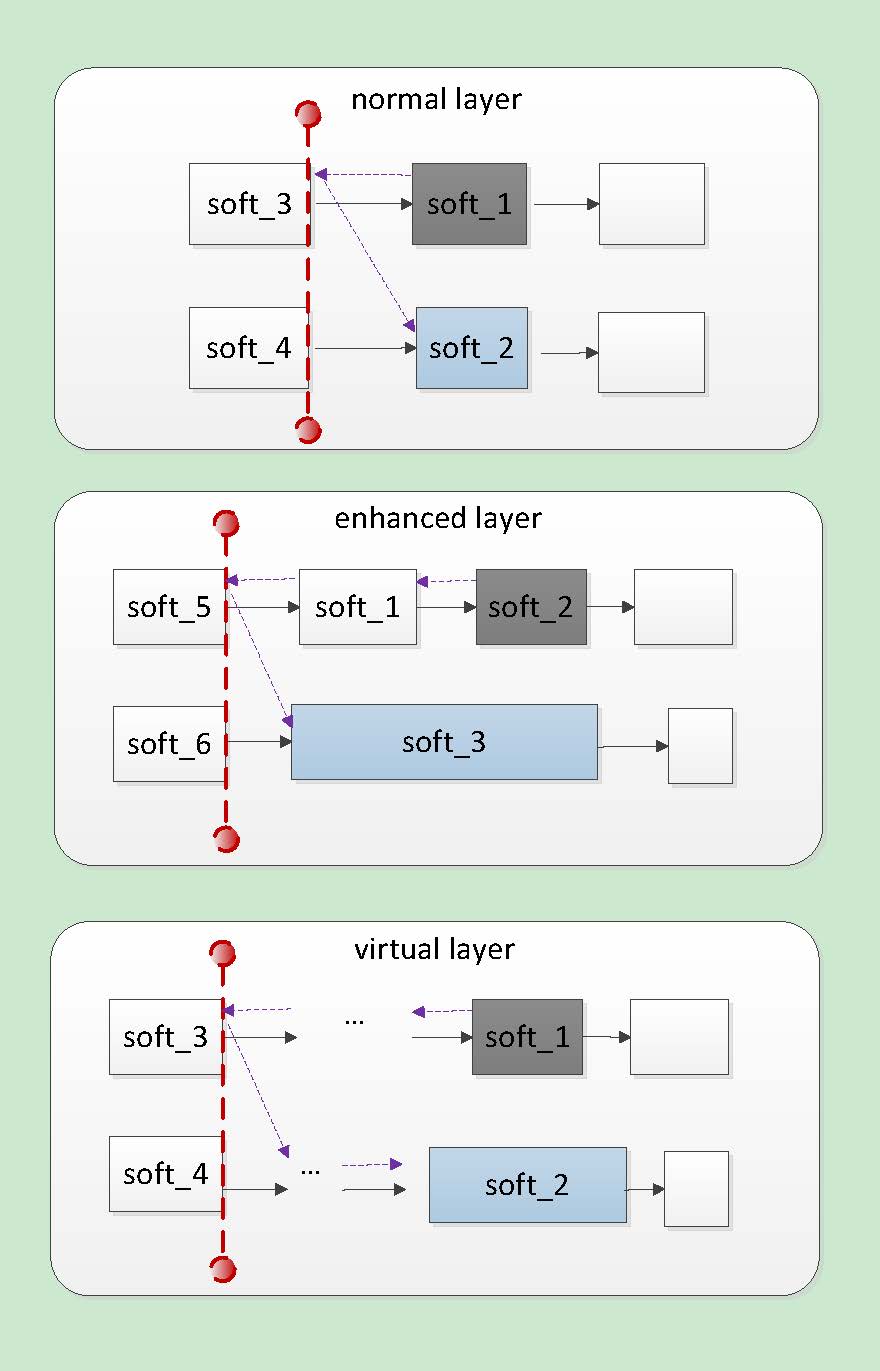}
    \caption{Switch path of different layers. The dotted arrow in purple color is the switch path.}
    \label{layer_retrey_path}
\end{figure}

Figure \ref{layer_retrey_path} shows the switching paths for different layers. Correspondingly, $len_{switch}$ contains the length of two steps, $len_{back}$ (the length of the back path) and $len_{forward}$ (the length of the forward path), as shown in \eqref {eq_len_recursive}.

\begin{equation} \label{eq_len_recursive}
  len_{switch} = len_{back} + len_{forward}
\end{equation}

Now, we begin to analyze the switch paths of these three types of layers.
The corresponding switch path is shown in Figure \ref{layer_retrey_path}.
The top of Figure \ref{layer_retrey_path} (normal layer) shows the switch path of the normal layer.
$len_{switch}$ is three because three pieces of software are involved, i.e. processing goes from $soft\_1$ back to $soft\_3$ and then switches to $soft\_2$.

The switch path of the enhanced layer is shown in the middle part of Figure \ref{layer_retrey_path}, where $len_{switch}$ is 4. In general, the $len_{switch}$ of an enhanced layer is $n$ + 2, where $n$ is the number of combined software aligned with other software. In Figure \ref{layer_retrey_path}, two software ($soft\_1$ and $soft\_2$) are combined to align with $soft\_3$, so $n$ is 2.

The switch path of the virtual layer is shown in the lower part of Figure \ref{layer_retrey_path}, and the minimum value of $len_{switch}$ for this figure is 5. Similarly, $len_{switch}$ is $k$+$l$+3, where $k$ and $l$ are the number of software in the ellipsis in the figure. The value of $k$ and $l$ is not less than 1, because there is at least one software under the corresponding path.

\subsubsection{P2P Processing Protocol Between Layers} 
P2P software peers interact through protocols between layers. A new P2P software peer follows the same protocol to interact with others when joining a working path. This protocol, known as the P2P Processing Protocol, enables seamless task switching and interaction between software peers.

The protocol is intended for use between software peers and does not necessitate a global protocol or the involvement of any centralized or individual software, with the aim of reducing dependence on centralized software. This type of agreement is called as a P2P agreement.

\begin{figure}[htb]
    \includegraphics[width = 3.5in]{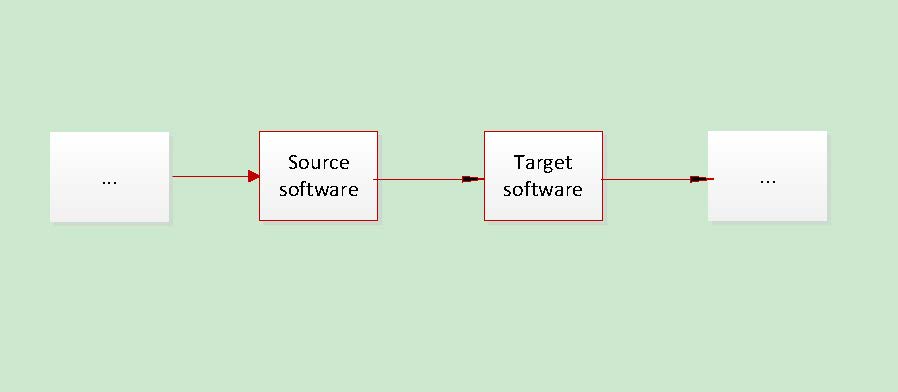}
    \caption{An example of source and target software peers}
    \label{target_soruce}
\end{figure}

If one software peer handles the subsequent task of another software peer, the former is called \textbf{target software} or \textbf{target software peer} and the latter is called \textbf{source software} or \textbf{source software peer}. The process is that the source software peer sends the processed task to the target software peer for further processing. An example is shown in Figure \ref{target_soruce}. If the source software peer is in layer $m$, its target software peer is in layer $n$, where $n > m$.

The task processing result should be returned to the preceding software (a source software peer or a terminal). After the target software peer finishes processing the task, it returns the processing result to the source software peer. This result is called a \textbf{return handle}. When the final task processing is completed, the final processing software peer returns the entire processing path to the terminal, called a \textbf{return path}. The terminal can use the return path to query the processing results in any software peer on the working path.

Here we discuss the terminal. The terminal is a system that sends task requests to P2P applications, which can be mobile phones or servers. The path to the P2P layer is configured in the terminal. After final processing, the return path is sent back to the terminal. 

\subsection{Working Procedure}
The processing flow of P2P software is from the terminal to the final processing software.
Assume that the terminal is $t_k$, and there are $n$ software peers (named from $soft^1_1$ to $soft^1_n$) in layer 1 ($layer_1$). The process is divided into the following three stages.

(1) Initial stage. At the beginning, $t_k$ chooses a software peer in $layer_1$ (suppose $soft^1_i$). If $soft^1_i$ does not respond within the specified time, it is considered down, and $t_k$ will choose another software peer from $layer_1$ until (1a) an available software peer($soft^1_j$) is found or ( 1b) all software peers in $layer_1$ do not work. If (1b) it will return an error return handle to $t_k$, which means that none of the peer to peer software cannot handle the task successfully and more P2P software peers required.

(2) Middle choice stage. Assuming that the current process is processed in software $soft^m_n$ at layer $layer_m$, the task processing will be passed from $layer_m$ to $layer_{m+1}$. There are three possible outcomes. (2a) If an available software peer ($soft^{m+1}_l$) is found, processing is passed to $layer_{m+1}$. (2b) If no software peer in $layer_{m+1}$ is found, $soft^m_n$ will notify the source software peer in $layer_{m-1}$ and retries in $layer_{m-1} $. (2c) If $soft^m_n$ becomes unresponsive due to a crash, its source software peer detects this and retries.

(3) Completion stage. A task is considered complete when the processing of the last layer is complete. In this case, the return path is given to terminal $t_k$.
If the feedback does not reach the terminal $t_k$ within the maximum time ($t^{require}_{max}$), the terminal must reissue the request, even if the process has been completed at the final layer.

\subsection{Key Measurements} \label{sec_key_measurement}
In this section, we describe two key measurements of P2P software: the successful probability and the task time. (1)The goal of P2P software is to increase the probability of task completion, so the first measurement is the degree of success of task completion, the successful probability.
(2) Different paths will affect the switching mechanism when a P2P software peer fails, resulting in different completion times. In order to measure the completion time, we introduce another measurement from the perspective of time, the average time for task completion or failure, named as the task time.

\subsubsection{Successful Probability}
The successful probability is the probability of task completion, denoted as $P_{success}$.
Since there are different paths to complete the task, the successful probability is the union of the success probabilities of all paths, as shown in \eqref{pro_task_success}.

\begin{equation} \label{pro_task_success}
    \begin{array}{ll}
       P_{success} = P_{success}(P_{path_1} \cup ... \cup P_{path_i} \cup ... \cup P_{path_n}) = \\
         \ \ \ \ \ \ \ \ \sum_{i=1}^{n}{(P_{path_i})} \\
         \ \ \ \ \ \ \ \ - \sum_{i=1}^{n-1}{(P_{path_i} P_{path_{i+1}})} \\
         \ \ \ \ \ \ \ \ + \sum_{i=1}^{n-2}{(P_{path_i} P_{path_{i+1}} P_{path_{i+2}})} \\
         \ \ \ \ \ \ \ \ - \sum_{i=1}^{n-3}{(P_{path_i} P_{path_{i+1}} P_{path_{i+3}} P_{path_{i+4}})} \\
         \ \ \ \ \ \ \ \ + ... \\
         \ \ \ \ \ \ \ \  - ...
    \end{array} 
\end{equation}

From \eqref{pro_task_success}, we can see that $P_{success}$ increases when the number of paths increases. This can be understood as when some paths fail, there are more alternative independent paths available to execute tasks.

Now we analyze the successful probability for different types of layers, normal, enhanced and visible layers. For simplicity, we only analyze the case where there are only two paths, and where each software peer has the same successful probability.

\begin{figure}[htb]
      \includegraphics[width=3.5in]{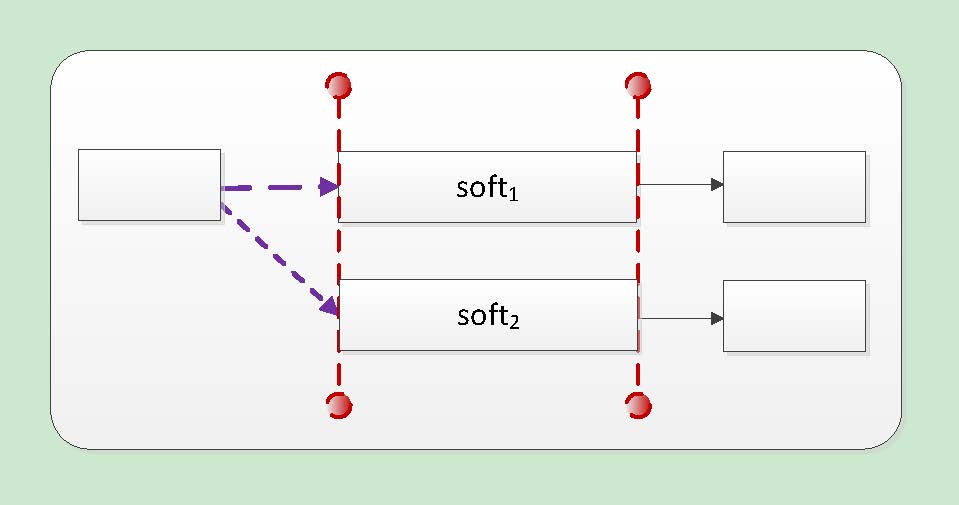}
      \caption{Normal layer. The red line indicates the same protocol (the same below). }
      \label{normal_layer_calculation}
\end{figure}

In the case of normal layers, the switch path can be switched at the same layer, as illustrated in Figure \ref{normal_layer_calculation}. Therefore, $P_{success}$ is shown in \eqref{pro_succ_normal}.

\begin{equation} \label{pro_succ_normal}
    P_{success} = P_{path_1} \cup P_{path_2}
\end{equation}

\begin{figure}[htb]
      \includegraphics[width=3.5in]{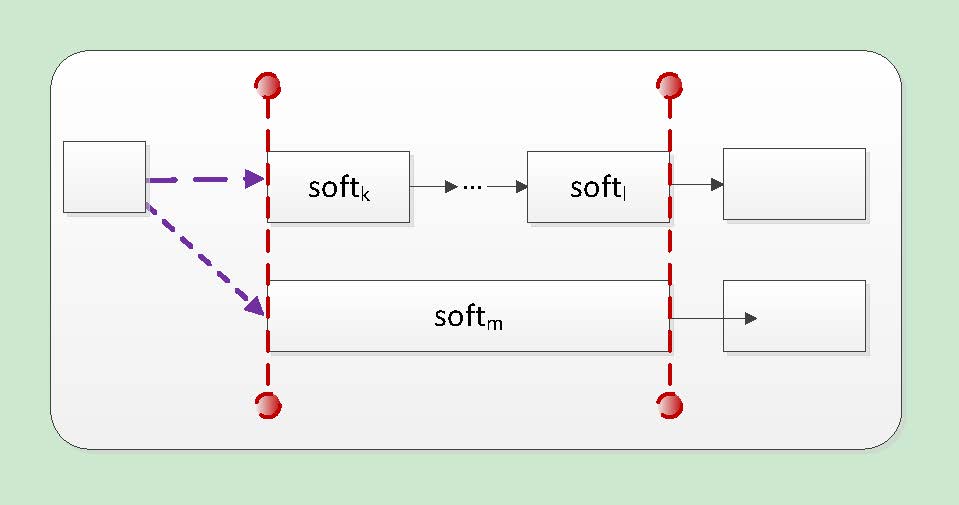}
      \caption{Enhanced layer. }
      \label{enhanced_layer_calculation}
\end{figure}

The successful probability of enhanced layers is often smaller than that of the normal layers. An example is shown in Figure \ref{enhanced_layer_calculation}, where the upper layer working path has multiple software peers to align with the lower layer working path. The corresponding $P_{success}$ is shown in \eqref{pro_succ_enhanced}. As can be seen from \eqref{pro_succ_enhanced}, the corresponding path has a smaller successful probability, since all software peers along that path must succeed for the task to complete. Therefore, it results in an overall smaller successful probability.

\begin{equation} \label{pro_succ_enhanced}
      P_{success} = (P_{soft_k} \cap ... \cap P_{soft_l}) \cup P_{soft_m}
\end{equation}

\begin{figure}[htb]
    \includegraphics[width=3.5in]{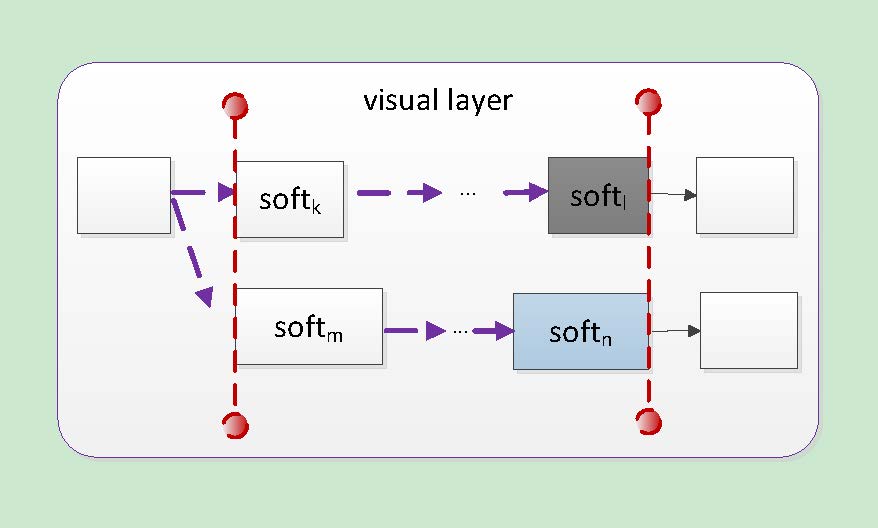}
    \caption{Visual layers.}
    \label{visual_layer_calculation}
\end{figure}

The virtual layer is shown in Figure \ref{visual_layer_calculation}. The corresponding successful probability is shown in \eqref{success_probability_visual}. There are more than one software peer involved in the back and forward paths, which results in the lowest successful probability of the three.

\begin{equation} \label{success_probability_visual}
    P_{success} = (P_{soft_{k}} \cap ... \cap P_{soft_{l}}) \cup (P_{soft_m} \cap ... \cap P_{soft_{n}})
\end{equation}

\subsubsection{Task time}
Task time refers to the average duration required to process a given task, from the moment it is sent by the terminal to the point at which task processing is completed. The completion of a task means that either the task has been successfully processed, or all available paths have been attempted and deemed unsuccessful.

Task time has an impact on the wait time required for the terminal, as per the P2P software model. Specifically, the terminal will either re-issue a task if a path fails, or record the return path if it succeeds. In general, longer task times require longer wait times for the terminal.

Now, we begin to analyze the maximum waiting time for a task to complete. The task time represents the average time required to complete a task. Some tasks may take longer than the task time to complete. Therefore, it is recommended for a terminal to wait for a sufficient amount of time before timing out. We set a maximum waiting time in each layer (denoted as $t^{i}_{max}$ for layer $i$). The sum $t^{total}_{max}$ for each layer is shown in \eqref{time_max_total_def}.

\begin{equation} \label{time_max_total_def}
      t^{total}_{max} = \sum_{1}^{n}{t^{i}_{max}}
\end{equation}
, where $n$ is the total number of layers.
\\

When the number of layers increases, $t^{total}_{max}$ may become larger. To reduce the impact of unknown latencies, we introduce a cache layer, named as the P2P cache layer. An P2P cache layer is the layer that caches task requests, and gives the return handle to the terminal when the target task completes.
When tasks are cached, terminal retries can be done by the caching layer. User terminals can even be turned off to save power. However, the terminal should send tasks to multiple software peers in the caching layer to avoid failure of one cache software peer.

When caching layers are introduced, the waiting time of a terminal, $t^{total}_{waiting}$, is displayed in \eqref{time_waiting_total_def}, where $k$ is the layer number of the caching layer. From \eqref{time_waiting_total_def}, we can see that when $k$ is smaller (closer to the terminal), $t^{total}_{waiting}$ is smaller. Therefore, the caching layer is usually positioned in layer 1, and the terminal only needs to wait for the request to be sent to the P2P cache layer.
 
\begin{equation} \label{time_waiting_total_def}
  t^{total}_{waiting} = \sum_{1}^{k}{t^{i}_{max}}
\end{equation}
, where $k$ is the layer number of the cache layer and $k$ \textless $n$.
\\

\section{Topology}
A working path is a linear structure, while multiple working paths may form a network structure, which we call the topology of a P2P application. Topology specifies the interaction relationship of P2P software peers.

To depict the network formed by working paths, a topology graph is employed. The topology graph comprises P2P software peers as vertices and the interaction relationships between P2P software peers as edges. The edges are directional, originating from the source software peer and terminating at the target software peer.
Correspondingly, a working path is the path (in topology) from a user request to the final processing software, representing the entire processing of a task.

\subsection{Separate and Coupled Paths}
\begin{figure}[htb]
  \includegraphics[width=3.5in]{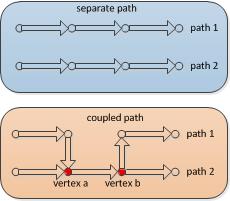}
  \caption{Separate path and coupled path. In the coupled path, there are two vertexes (vertex a and b in red) which are in path 1 and path 2; while in the separate paths, there is no such vertexes.}
  \label{separate_path}
\end{figure}

Regarding whether the paths have the same software peer, the paths can be divided into separate paths and coupled paths. Separate paths are paths that have no common vertices, it is also called \textbf{basic path}. Meanwhile, paths with common vertices are called coupled paths. Figure \ref{separate_path} shows separated and coupled paths.

Both separate and coupled paths can be working paths because they contain enough software peers required to complete a task. Therefore, the number of working paths is the sum of separate paths and coupled paths. We use a variable $Path_{task}$ to represent the collection of all working paths, as shown in \eqref{path_collection_definition}.

\begin{equation} \label{path_collection_definition}
      Path_{task} = \{ path_{wk_1}, ..., path_{wk_n} \}
    \end{equation}
    , where $path_{wk_i}$ is a separate or coupled path.
\\

If each software peer on a working path has an equal successful probability, then a separate path will have a higher successful probability than a coupled one. Assume that there are two paths with success probabilities $p1$ and $p2$, and the final successful probability is shown in \eqref{p_success_separate_coherent}. The value of $p1*p2$ is 0 in the separate path, because they have no common software peer, so the successful probability of the separate path is greater.

\begin{equation} \label{p_success_separate_coherent}
  p_{success} = p1 + p2 - p1*p2
\end{equation}

% \begin{figure}[htb]
%   \includegraphics[width=3.5in]{separate_path_number}
%   \caption{The topology in the left contains two separate paths and three coherent paths.}
%   \label{separate_path_number}
% \end{figure}

\subsection{Topology Types}
It's not always possible to optimize both the successful probability and task time simultaneously. If we want to increase the successful probability, we may need to add more working paths. However, more working paths can also lead to more switching, which increases the task time. Therefore, different topologies offer the opportunity to choose a trade-off between these two measurements.

\subsubsection{Full Connection}
\begin{figure}
    \includegraphics[width=3.5in]{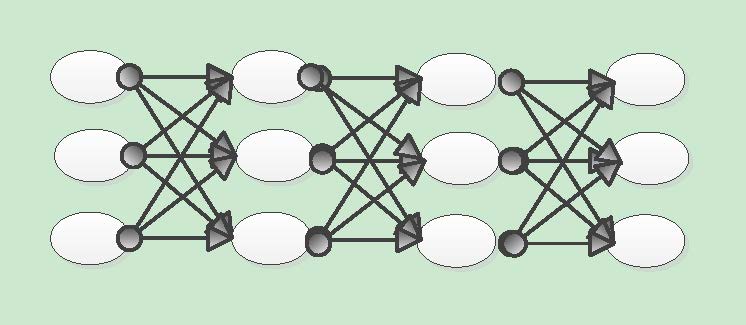}
    \caption{Example of full connection. All possible edges within the same layer are connected, no more edges can be added. }
    \label{full_connection}
\end{figure}

Full connection is a P2P topology, where each software peer on the previous layer connects to all software peers on the next layer.
In the corresponding topological graph, all possible vertices within the same layer are connected and no more edges can be added, as shown in Figure \ref{full_connection}.

Correspondingly, the successful probability is maximized for full connections, since all possible retry paths can be used. The task can only fail if there are no available paths from the source software peer to the target software peers, as in \eqref{path_relation_in_full_connection}.

\begin{equation} \label{path_relation_in_full_connection}
    \forall path \in Path_{task}, path \in Path_{full}
\end{equation}
, where $Path_{full}$ is all paths in the full connection.
\\

On the other hand, in the full connection, the task time is maximal. In the worst case, all possible paths to the next layer will be retried, as shown in \eqref{retry_time_full}.

\begin{equation} \label{retry_time_full}
  time^{retry\_full}_{max} = \sum^{path_{i} \in Path_{full}}{time^{retry}_{path_{i}}}
\end{equation}
, where $time^{retry\_full}_{max}$ is the maximum task time of full connection, and $time^{retry}_{path_{i}}$ is the task time taken by $path_{i}$.
\\

\subsubsection{Minimum Connection}
In some cases, it's necessary to limit the task time, while ensuring a high successful probability. We can remove some paths to reduce retries on working paths. Meanwhile, a software peer should connect to at least $m$ software peers to ensure that $P_{success} \textgreater \delta$. Although, the minimum connection does not cover all possible paths,  it guarantees a successful probability greater than $\delta$.

\begin{equation} \label{path_relation_in_minimum_connection}
    \begin{array}{l}
      Path_{minimum} \subset Path_{task} \\
      P_{success}(Path_{minimum}) > \delta
    \end{array}
\end{equation}
, where $Path_{minimum}$ is the set of all paths in the minimum connection.
\\

From \eqref{path_relation_in_minimum_connection}, we see that $P_{success}(Path_{minimum})$ is less than $P_{success}(Path_{full})$ because $Path_{minimum} \in Path_{full}$ .

The task time of the minimum connection is the sum of the attempt times of its paths, as shown in \eqref{retry_time_minimum}, which is less than the task time of the full connection, as shown in \eqref{retry_time_full}.

\begin{equation} \label{retry_time_minimum}
      time^{retry\_minimum}_{max} = \sum^{path_{i} \in Path_{minimum}}{time^{retry}_{path_{i}}}
  \end{equation}
  , where $time^{retry\_minimum}_{max}$ is the maximum task time of the minimum connection, and $time^{retry}_{path_{i}}$ is the task time occupied by $path_{i}$.

Comparing the minimum connection and the full connection, there is a trade off between them: although the successful probability of the minimum connection is smaller, its maximum task time is smaller.
\subsection{Add New Software Peer to Topology}
When a new software peer is added, its successful probability is uncertain, although it may have been verified elsewhere.
The successful probability in the new environment should be measured by the actual processing results thereafter.

Since a new software peer does not guarantee a successful probability, the software peer should not be relied upon to complete tasks. Otherwise, there may be situations where the introduction of new software fails (as in COVID-19).

Therefore, the method of \textbf{double sending} is used. The double sending method is to send the request to multiple target software peers. In this case, the request is sent to a known software peer and another to the new software peer. Tasks are processed in parallel. 
If a task is successfully processed by the new software within a certain amount of time, the software's successful probability increases; otherwise, the successful probability decreases. 
This paper proposes to keep the successful probability software in its source software because (1) this eliminates a centralized software to record the successful probability, and (2) it is the source software that selects the target software.

When the successful probability of a new software peer reaches a certain value, the double sending can no longer be used.
And double sending can also be used to save task time. If a task fails to be processed in one working path, there is still another working path, which saves the corresponding retry and waiting time.

% <adding a working path or a component ??>

\section{Verification}
\subsection{Test Environment}
The test environment are described in three aspects: the P2P software, the topology formation, and the collection of corresponding measurements.

The whole system is composed of different P2P software peers, each of which is assigned an independent successful probability. In this verification, each P2P software is a web server, obtaining and processing task requests from other P2P software peers.
After a P2P software peer finishes processing a task, the P2P software peer further transfers the task to its target software peer by calling the corresponding web interface.

The simulation of being down is as follows. When a P2P software peer gets a task, it generates a random number in a certain range (0 to N, N is sys.maxsize in python, 9223372036854775807 in the verification system). If it is less than ($1 - p$)*N , the software peer is set to hang and simulated by sleeping infinitely.
Otherwise, the task processing succeeds: (1) the signature of the processing software is added to the task's data field, and (2) to simulate the processing time of the task, it sleeps for a few seconds. 
%Then a return handle is given to the source software peer.  
If the source software does not get the return handle within a certain period of time, it is considered that the target is down.

The P2P topology information is configured into each software peer, without the need for a centralized software to maintain such information. One of the important information is the interface parameters of the target software peer. This information is configured in its source software. When the target software is down, the source software can select another target software through this information.

The result of the task is judged as follows. Judgment of task success: the final processing software completes the processing and passes the corresponding return handle to the terminal (because the cache layer is not used during verification, the return handle should be returned to the terminal). Judgment of task failure: all paths have been tried and task processing failed or no return handle has been provided to the terminal layer.

The corresponding measurements are collected below. For the successful probability, we use two counters, $counterS$ and $counterF$.
When the task succeeds, $counterS$ is incremented by 1, and when the task fails, $counterF$ is incremented by 1. The successful probability is calculated as: $p_{success} = counterS / (counterS + counterF)$.
For task time, two related time indicators are recorded: (1) the start time is recorded when the terminal sends the task; (2) the final time is recorded when the task fails or succeeds.
The task time is the difference between the end time and the start time.

\subsection{Successful Probability} \label{sec_suc_pro}
In this section, we try to verify the factors that affect the successful probability, including the impact of different paths, the impact of coupling, and the impact of division granularity.

\subsubsection{Impact of Different Path}
To form the different paths, we used several topologies: different numbers of basic paths and different ways of connecting them.
(1) The types of basic paths are 2-basic paths, 3-basic paths, and 4-basic paths.
(2) Connections between different layers include: separate paths, fully connected layers 2 (layer 2 full), fully connected layers 2 and 3 (layer 2, 3 full), and fully connected layers 2, 3, and 4 (ayer 2, 3 and 4 full). Separate paths mean that all basic paths are independent, i.e. software in one basic path is not connected to software in other paths. 
A fully connected layer means that every piece of software in the previous layer is connected to all software in that layer.

\begin{figure}[htp]
  \includegraphics[width=3.5in]{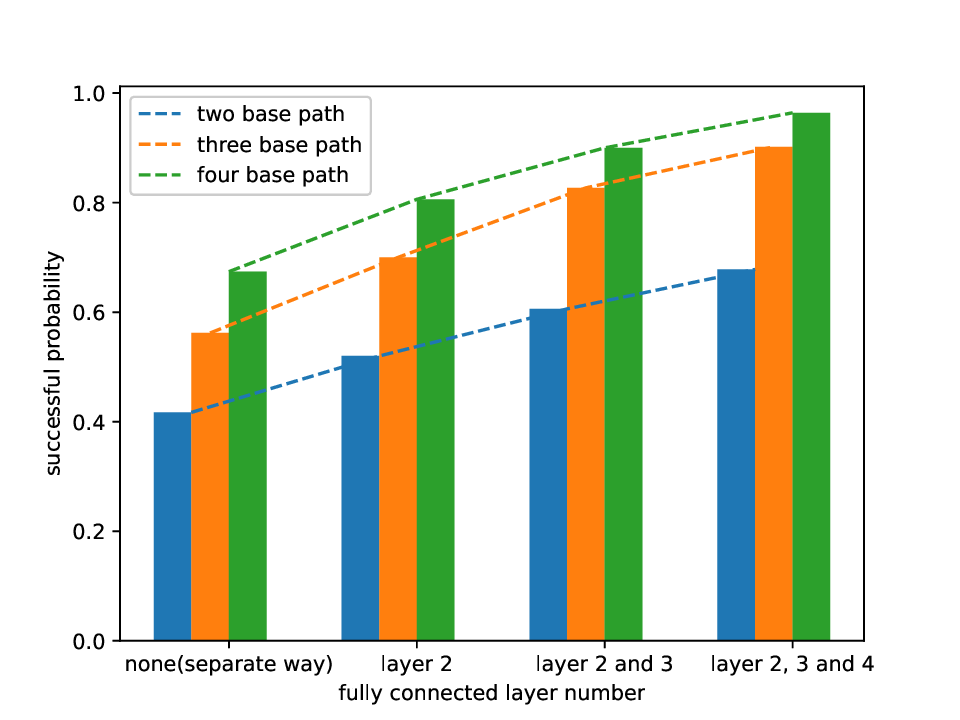}
  \caption{The impact of different path number.}
  \label{pathNumberImpact}
\end{figure}

After combining the basic paths and connection methods, there are 12 test scenarios in total. Each scenario was executed 1000 times. The result is shown in Figure \ref{pathNumberImpact}, which shows two interesting points.
(1) When the number of basic paths increases, the successful probability increases. Taking layer 2 full as an example, when the number of basic paths increases from 2 basic paths to 4 basic paths, the successful probabilities are 0.52, 0.70, and 0.81, respectively.
(2) When the number of fully connected layers increases, the successful probability increases. Considering 3 basic scenarios, the successful probabilities are 0.56, 0.70, 0.83, 0.90 seconds respectively.
This suggests that if we want to increase the successful probability, we either increase the number of basic paths or the number of full connections.

\subsubsection{Impact of Coupling Path}
We choose two basic paths with four layers for verification, and there are four cases depending on the software used in common, named 0common, 1common, 2common and 3common respectively. The first number in these names indicates the amount of the same software peer(s) used. For instance, 2common indicates that these paths have two identical software peers. 0common means that these paths do not have any identical peers, i.e. they are separate paths.
Using identical software peers creates interdependence between the paths. If there's a failure in the commonly used software peer, both paths will fail simultaneously, i.e. these paths are coupled by the identical software peer.

\begin{figure}[htp]
    \includegraphics[width=3.5in]{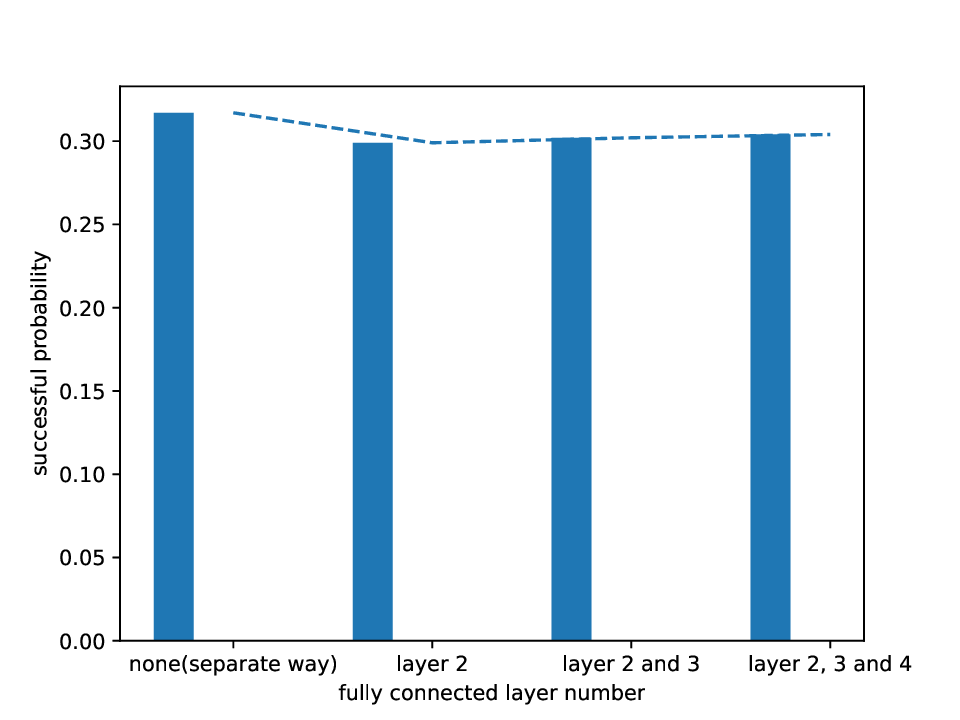}
    \caption{The successful probability affected by coupled path number. }
    \label{coherenetImpact}
\end{figure}

Each scenario is executed 1000 times and the results are shown in Figure \ref{coherenetImpact}. It can be seen from the figure that the successful probability is the higher when all paths have no common software (none, separate way), 0.32. The successful probabilities of the other three cases are lower, 0.299, 0.302, 0.304 respectively. The verification results show that separate paths have a greater successful probability.

\begin{figure}[htp]
    \includegraphics[width=3.5in]{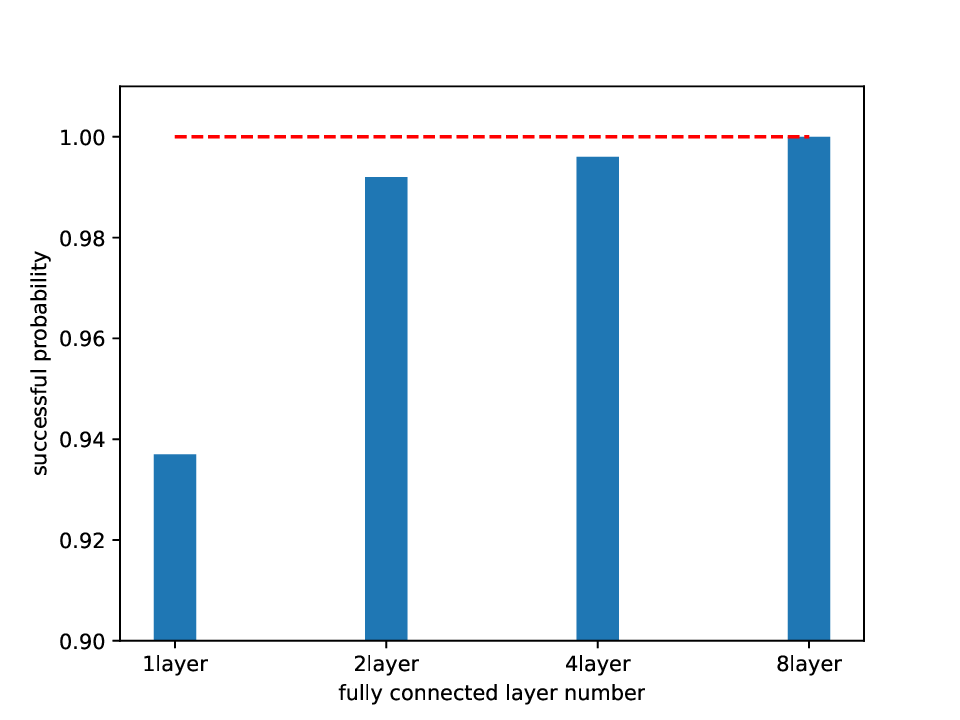}
    \caption{Impact of layer division on the successful probability. }
    \label{divisionSizeImpact}
\end{figure}

% Move task time here to adjust the figures' postion
\begin{figure*}[htb]
    \includegraphics[width=7in]{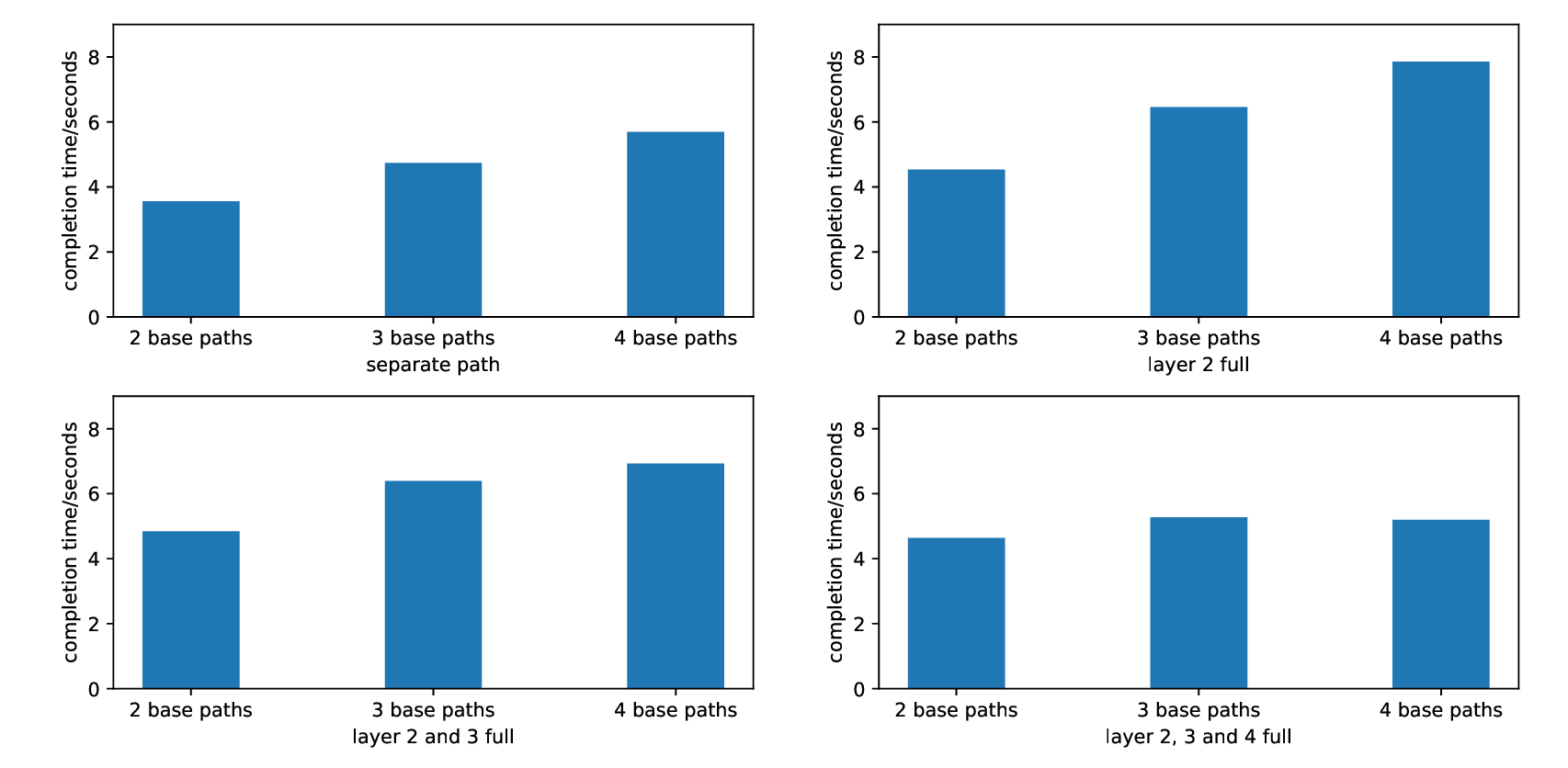}
    \caption{The task time affected by different path number.}
    \label{pathNumberCompletionTime}
  \end{figure*}
  
  \begin{figure}[htp]
    \includegraphics[width=3.5in]{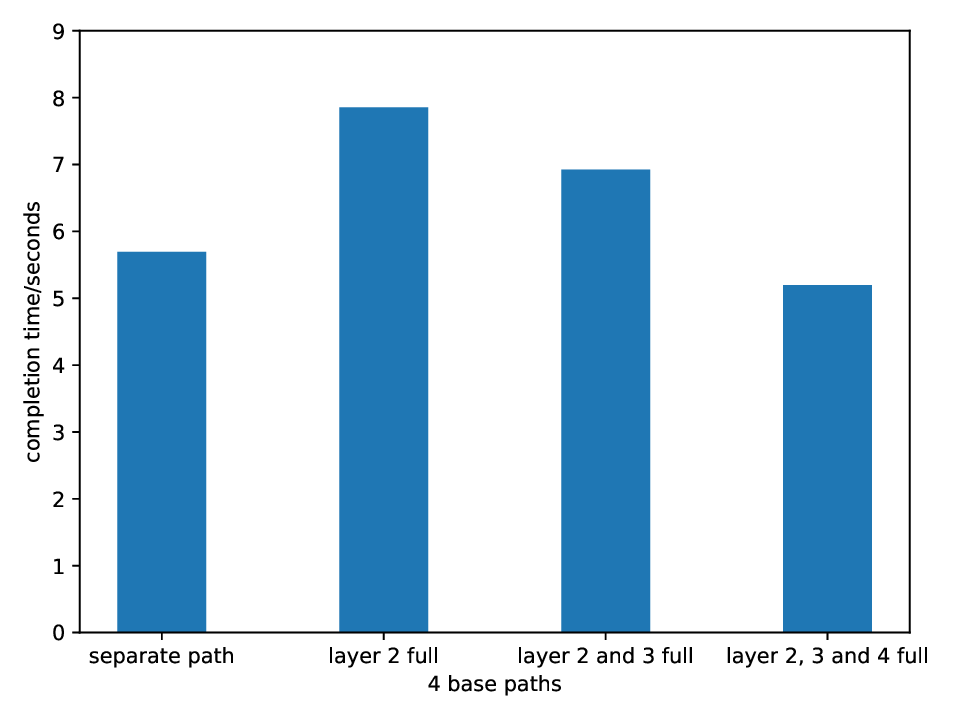}
    \caption{The task time affected by full connection with 4 basic paths.}
    \label{retryTime4BasePath}
  \end{figure}

\subsubsection{Impact of Division Granularity} \label{division_size_success_prob_ver_sec}
In order to verify the impact of division granularity on the successful probability, we designed four types of verification: (1) there is no division in the basic path, named as 1layer, (2) each basic path is divided into 2 layers, named as 2layer, (3 ) each basic path is divided into 4 layers, named as 4layer, and (4) each basic path is divided into 8 layers, named as 8layer.

Layer division has the following restrictions.

(1) The connections between the basic paths are fully connected.
The successful probability for the basic path is set to 0.5. The reason is that if the successful probability of a basic path is set to 0.7, then when all paths are fully connected, the successful probability will reach 99.1\% ($1 - (1-0.7)^4$), which makes the difference between different verification scenarios smaller.

(2) The successful probability ($p_{success}^i$) of a task is calculated by the successful probability $p_{success}$ of the basic path.
If the basic path is divided into $n$ layers. Their total successful probability is equal to $p_{success}$, as shown by \eqref{success_probability_layer_division}.

\begin{equation}\label{success_probability_layer_division}
  p_{success} = p_{success}^1 * p_{success}^2 * ... * p_{success}^n
\end{equation}
, where $p_{success}^i$ is the successful probability of the $i$th software in a basic path.
In the verification, the successful probability is the same for each software, so we get $p_{success}^i = \sqrt[n]{p_{success}}$.
\\

Each scenario was tested 1000 times, and the results are shown in Figure \ref{divisionSizeImpact}. The successful probabilities of layers 1, 2, 4, and 8 are 0.937, 0.992, 0.996, and 1.0, respectively. The successful probability increases by 0.057 from the case of no layer division (1layer) to 2 layers (2layer), while the increase from 2layer to 4layer is only 0.004 because the successful probability is close to 1.0.
This verification shows that more layer divisions increase the successful probability

\subsection{Task Time}
In this section, we want to verify the factors that affect the task time. There are four sub-verifications, the impact of different paths, the impact of coupling, the image of division granularity and the impact of double sending.

\subsubsection{Impact of Different Paths}
In this verification, we verify the impact of different paths on the task time, including topologies with different number of paths and different connection methods.
The test scenario is still 12 different basic paths (2-basic path, 3-basic path, 4-basic path) and different connection methods (separate, layer 2 full, layer 2 and 3 full, and layer 2 3 and 4 full).

Each scenario was executed 1000 times. The result is shown in Figure \ref{pathNumberCompletionTime} and Figure \ref{retryTime4BasePath}. Figure \ref{pathNumberCompletionTime} shows the task time comparison between different numbers of paths, from which we can see that the task time increases as the number of basic paths increases. This is because when the number of basic paths increases, more paths can be retried, resulting in more task time.

Figure \ref{retryTime4BasePath} shows the task time comparison between different connection methods. We show the task times for the 4 basic path cases because the other two basic path cases have similar trends. The task time increases first and then decreases, which are 5.70, 7.86, 6.92, and 5.20 seconds respectively.
The increase in task time is due to the increase in the number of paths that can be retried. The subsequent decrease in task time is due to the increased successful probability: the probability that the software fails at a layer is only 0.0081 ($(1 - 0.7)^4$), because more paths can be retried. A task can be executed successfully in this layer without returning to the previous layer.
For example, the case of 2, 3, and 4 fully-connected layers has shorter task times than the separate path.

\subsubsection{Impact of Coupling}
\begin{figure}[htp]
    \includegraphics[width=3.5in]{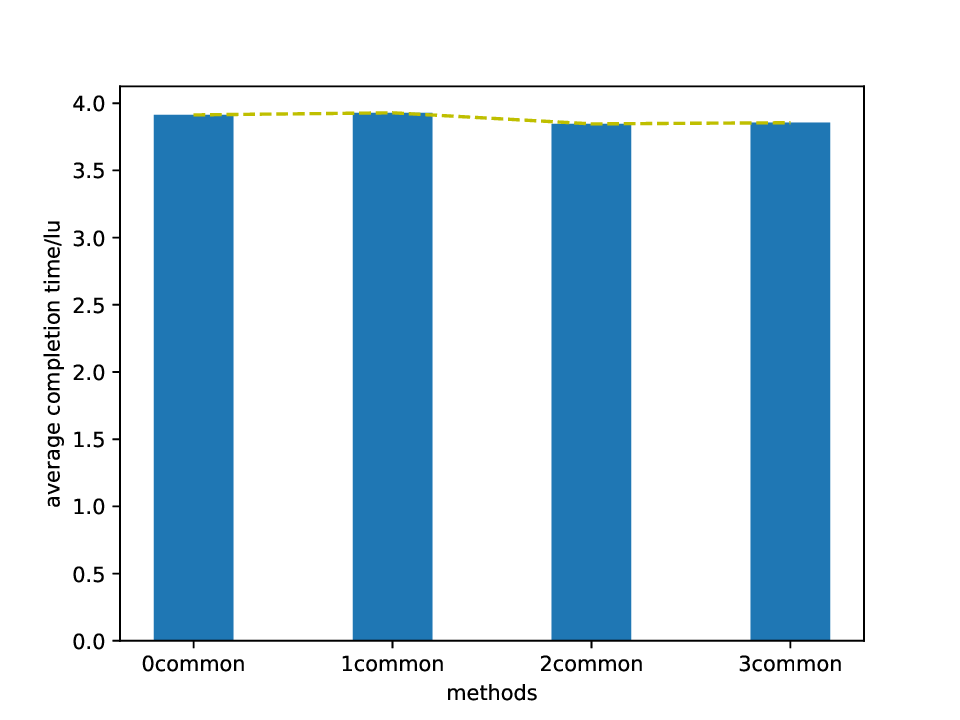}
    \caption{Impact of coupled paths on average task time. }
    \label{completionTimeBycoherentImpact}
\end{figure}

In this part, we verify the impact of software coupling on task timing. Verification scenarios include scenarios using two paths with four layers (0common, 1common, 2common, and 3common, similar to the scenarios in section \ref{sec_suc_pro}). Each scenario is tested 1000 times, and we take the average task time, as shown in \ref{completionTimeBycoherentImpact}.

From Figure \ref{completionTimeBycoherentImpact}, it can be seen that there is not much difference between these cases, and the average task time for the cases from 0common to 3common is 3.91, 3.93, 3.85, 3.85 seconds, respectively. The difference between the maximum and minimum is 0.08 seconds, which is only 2.04\% of the average task time of the maximum (1common).

\subsubsection{Impact of Layer Division Granularity}
\begin{figure}[htp]
    \includegraphics[width=3.5in]{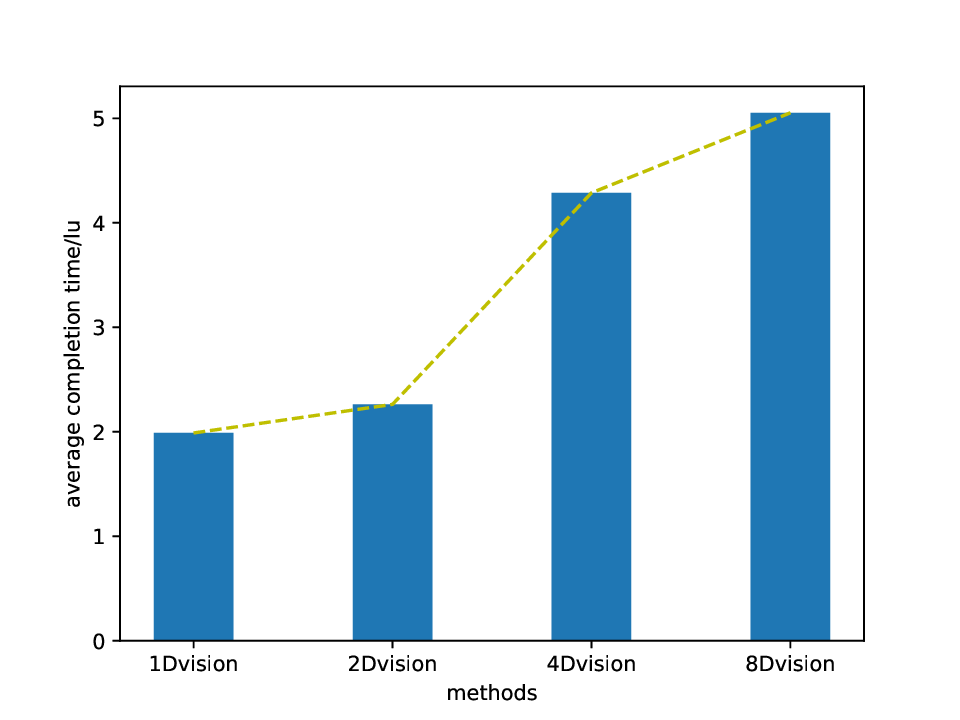}
    \caption{Impact of number of coupled paths on task time. }
    \label{completionTimedivisionSizeImpact}
\end{figure}

We still adopt the four verification scenarios in the \ref{division_size_success_prob_ver_sec} section, aiming to verify the impact of different layer division granularity on task time.
The verification was carried out 1000 times, and the average task time is shown in Figure \ref{completionTimedivisionSizeImpact}. From this figure, we can see that the task time increases by more than 2 seconds, from 2.26 seconds (2layer case) to 4.28 seconds (4layer case). This is because there are many paths that need to be retried during task processing.
While the task time from the 4layer case to the 8layer case only increased by 0.77 seconds, from 4.28 seconds to 5.05 seconds. The reason is that the successful probability is almost close to 1 at this time (the successful probability is from 0.996 to 1.0 in Figure \ref{divisionSizeImpact}), so that the task time does not increase significantly.

\subsubsection{Double Sending}
In this verification, we verified the impact of double sending on task time, using two scenarios named noDoubleSending and doubleSending respectively. We adopt a full connection with 4 basic paths and set the successful probability of each basic path to 0.5.
If there are two pieces of software in the subsequent layer that have not processed a task, the task is sent to the two subsequent pieces of software for processing. A task is considered successful if there is one final piece of software that successfully processes the task and gives the return handle to the terminal. If none of the software can handle the task, task processing is considered as failed.

\begin{figure}[htp]
  \includegraphics[width=3.5in]{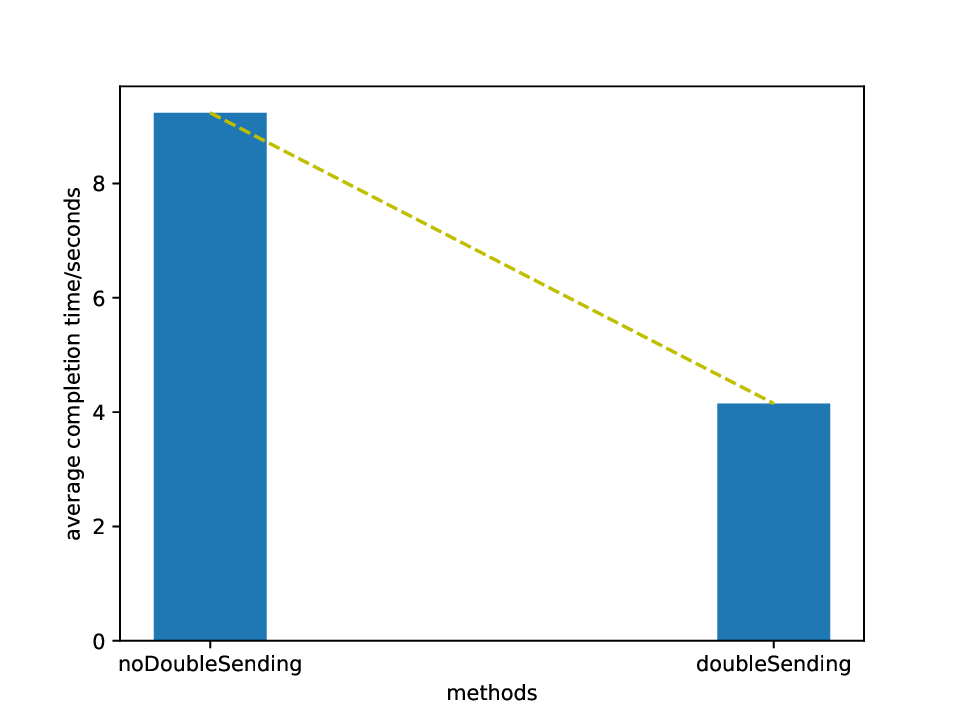}
  \caption{Average task time affected by double sending.}
  \label{doubleSending}
\end{figure}

The two scenarios are also carried out 1000 times, and the result is shown in Figure \ref{doubleSending}. As can be seen from the figure, the noDoubleSending task takes almost twice as long as the doubleSending method. Their average task times were 9.24 and 4.15 seconds, respectively.

This verification shows that double sending can save a lot of time (maybe only 50\% of the original task time).
But it also has some disadvantages: in the double sending method, a task may be successfully processed twice,
This can be a waste of computing resources.

\section{Conclusion}
% In this paper, we discuss how to achieve high availability for large-scale screening software. New screening software often needs to be deployed under tight deadlines, which can often result in downtime.
% To reduce the impact of the introduction of new software, we propose a P2P software model that allows new software and old software to form a topology with multiple independent processing paths.
% When one path (new software or old software) fails, there is another path to handle tasks, resulting in higher availability than a single software solution.
% Meanwhile, we also analyze different topologies that provide a balance between high availability and task time.
% The verification results show that the P2P software model with multiple processing paths can achieve a higher successful probability. Overall, we provide a P2P software approach to high availability of software used in large screening scenarios of highly contagious diseases (such as COVID-19).

This paper has presented a novel software in Peer-to-Peer (P2P) model designed to enhance high availability at the software level, addressing the critical gap in existing P2P architectures. By emphasizing the decentralized nature of software entities and advocating for independent, autonomous operation, the proposed model offers a promising solution for achieving robust software-level high availability. Moreover, the model obviates the need for software with central functions, such as task scheduling or dispatching. Through the establishment of a collaborative network topology with multiple autonomous processing paths, the model ensures seamless task transitions and continuous operation, even in the face of software-related disruptions or failures. 

The comprehensive analysis and validation of the proposed P2P software model have demonstrated its effectiveness in improving the probabilities of successful task processing while maintaining high availability. By overcoming the constraints of conventional redundancy methods and centralized software coordination, the model provides an adaptive and scalable framework that enhances the resilience and adaptability of P2P networks. The flexibility offered by the software peers in joining or leaving the network without relying on a centralized software infrastructure marks a advancement in the field, promising a more robust and flexible architecture for future system designs.

% use section* for acknowledgment
% \section*{Acknowledgment}
% The authors thank the anonymous reviewers for their constructive comments, which help us to improve the quality of this paper. This work was supported in part by the National Natural Science Foundation of China under Grant No. 61772352; the Science and Technology Planning Project of Sichuan Province under Grant No. 2019YFG0400, 2018GZDZX0031, 2018GZDZX0004, 2017GZDZX0003, 2018JY0182, 19ZDYF1286.

% Can use something like this to put references on a page
% by themselves when using endfloat and the captionsoff option.
\ifCLASSOPTIONcaptionsoff
  \newpage
\fi

\end{document}